subm

\documentclass[12pt,a4paper]{article}

\usepackage{ifthen} 
\newboolean{pdflatex}
\setboolean{pdflatex}{true} 

\newboolean{articletitles}
\setboolean{articletitles}{true} 

\newboolean{uprightparticles}
\setboolean{uprightparticles}{false} 

\newboolean{inbibliography}
\setboolean{inbibliography}{false} 

\usepackage{microtype}
\usepackage{lineno}  
\usepackage{xspace} 
\usepackage{caption} 

\usepackage{graphicx}  
\usepackage{color}
\usepackage{colortbl}
\graphicspath{{./figs/}} 

\usepackage{amsmath} 
\usepackage{amssymb}
\usepackage{amsfonts}
\usepackage{upgreek} 

\newcommand*\patchAmsMathEnvironmentForLineno[1]{%
\expandafter\let\csname old#1\expandafter\endcsname\csname #1\endcsname
\expandafter\let\csname oldend#1\expandafter\endcsname\csname
end#1\endcsname
 \renewenvironment{#1}%
   {\linenomath\csname old#1\endcsname}%
   {\csname oldend#1\endcsname\endlinenomath}%
}
\newcommand*\patchBothAmsMathEnvironmentsForLineno[1]{%
  \patchAmsMathEnvironmentForLineno{#1}%
  \patchAmsMathEnvironmentForLineno{#1*}%
}
\AtBeginDocument{%
\patchBothAmsMathEnvironmentsForLineno{equation}%
\patchBothAmsMathEnvironmentsForLineno{align}%
\patchBothAmsMathEnvironmentsForLineno{flalign}%
\patchBothAmsMathEnvironmentsForLineno{alignat}%
\patchBothAmsMathEnvironmentsForLineno{gather}%
\patchBothAmsMathEnvironmentsForLineno{multline}%
\patchBothAmsMathEnvironmentsForLineno{eqnarray}%
}

\usepackage{hyperref}    
\usepackage[all]{hypcap} 

\def\lhcb {\mbox{LHCb}\xspace}

\def\MagUp {\mbox{\em Mag\kern -0.05em Up}\xspace}

\ifthenelse{\boolean{uprightparticles}}%
{

 \def\Ppi         {\ensuremath{\uppi}\xspace}

 \def\PDelta      {\ensuremath{\Delta}\xspace}                 
 \def\PXi      {\ensuremath{\Xi}\xspace}                 
 \def\PLambda      {\ensuremath{\Lambda}\xspace}                 
 \def\PSigma      {\ensuremath{\Sigma}\xspace}                 
 \def\POmega      {\ensuremath{\Omega}\xspace}                 
 \def\PUpsilon      {\ensuremath{\Upsilon}\xspace}                 
 
 \def\PB      {\ensuremath{\mathrm{B}}\xspace}                 
                  
 \def\PD      {\ensuremath{\mathrm{D}}\xspace}

 \def\PK      {\ensuremath{\mathrm{K}}\xspace}

 \def\Pb      {\ensuremath{\mathrm{b}}\xspace}                 
 \def\Pc      {\ensuremath{\mathrm{c}}\xspace}

 \def\Pi      {\ensuremath{\mathrm{i}}\xspace}

}
{

 \def\Ppi         {\ensuremath{\pi}\xspace}

 \mathchardef\PDelta="7101
 \mathchardef\PXi="7104
 \mathchardef\PLambda="7103
 \mathchardef\PSigma="7106
 \mathchardef\POmega="710A
 \mathchardef\PUpsilon="7107
                  
 \def\PB      {\ensuremath{B}\xspace}                 
                  
 \def\PD      {\ensuremath{D}\xspace}

 \def\PK      {\ensuremath{K}\xspace}

 \def\Pb      {\ensuremath{b}\xspace}                 
 \def\Pc      {\ensuremath{c}\xspace}

 \def\Pi      {\ensuremath{i}\xspace}

}

\makeatletter
\ifcase \@ptsize \relax
  \newcommand{\miniscule}{\@setfontsize\miniscule{4}{5}}
\or
  \newcommand{\miniscule}{\@setfontsize\miniscule{5}{6}}
\or
  \newcommand{\miniscule}{\@setfontsize\miniscule{5}{6}}
\fi
\makeatother

\DeclareRobustCommand{\optbar}[1]{\shortstack{{\miniscule (\rule[.5ex]{1.25em}{.18mm})}
  \\ [-.7ex] $#1$}}

\def\cquark    {{\ensuremath{\Pc}}\xspace}

\def\bquark    {{\ensuremath{\Pb}}\xspace}

\def\pion   {{\ensuremath{\Ppi}}\xspace}

\def\pip    {{\ensuremath{\pion^+}}\xspace}
\def\pim    {{\ensuremath{\pion^-}}\xspace}
\def\pipm   {{\ensuremath{\pion^\pm}}\xspace}

\def\kaon    {{\ensuremath{\PK}}\xspace}
  \def\Kbar    {{\kern 0.2em\overline{\kern -0.2em \PK}{}}\xspace}

\def\KorKbar    {\kern 0.18em\optbar{\kern -0.18em K}{}\xspace}

\def\Km      {{\ensuremath{\kaon^-}}\xspace}

  \def\Dbar    {{\kern 0.2em\overline{\kern -0.2em \PD}{}}\xspace}
\def\D       {{\ensuremath{\PD}}\xspace}

\def\DorDbar    {\kern 0.18em\optbar{\kern -0.18em D}{}\xspace}
\def\Dz      {{\ensuremath{\D^0}}\xspace}

\def\Dstarp  {{\ensuremath{\D^{*+}}}\xspace}

\def\Bbar    {{\ensuremath{\kern 0.18em\overline{\kern -0.18em \PB}{}}}\xspace}

\def\BorBbar    {\kern 0.18em\optbar{\kern -0.18em B}{}\xspace}

  \def\Y#1S{\ensuremath{\PUpsilon{(#1S)}}\xspace}

\def\Xires       {{\ensuremath{\PXi}}\xspace}

\def\Lz          {{\ensuremath{\PLambda}}\xspace}
\def\Lbar        {{\ensuremath{\kern 0.1em\overline{\kern -0.1em\PLambda}}}\xspace}
\def\LorLbar    {\kern 0.18em\optbar{\kern -0.18em \PLambda}{}\xspace}

\def\Sigmares    {{\ensuremath{\PSigma}}\xspace}

\def\Lb      {{\ensuremath{\Lz^0_\bquark}}\xspace}
\def\Lbbar   {{\ensuremath{\Lbar{}^0_\bquark}}\xspace}
\def\Lc      {{\ensuremath{\Lz^+_\cquark}}\xspace}

\def\Xibz    {{\ensuremath{\Xires^0_\bquark}}\xspace}
\def\Xibm    {{\ensuremath{\Xires^-_\bquark}}\xspace}

\def\to                 {\ensuremath{\rightarrow}\xspace}

\def\CP                {{\ensuremath{C\!P}}\xspace}

\def\AT#1     {\ensuremath{A_{\mathrm{T}}^{#1}}\xspace}           

\def\C#1      {\ensuremath{\mathcal{C}_{#1}}\xspace}                       
\def\Cp#1     {\ensuremath{\mathcal{C}_{#1}^{'}}\xspace}                    
\def\Ceff#1   {\ensuremath{\mathcal{C}_{#1}^{\mathrm{(eff)}}}\xspace}        
\def\Cpeff#1  {\ensuremath{\mathcal{C}_{#1}^{'\mathrm{(eff)}}}\xspace}       
\def\Ope#1    {\ensuremath{\mathcal{O}_{#1}}\xspace}                       
\def\Opep#1   {\ensuremath{\mathcal{O}_{#1}^{'}}\xspace}                    

\newcommand{\tev}{\ifthenelse{\boolean{inbibliography}}{\ensuremath{~T\kern -0.05em eV}\xspace}{\ensuremath{\mathrm{\,Te\kern -0.1em V}}}\xspace}
\newcommand{\gev}{\ensuremath{\mathrm{\,Ge\kern -0.1em V}}\xspace}
\newcommand{\mev}{\ensuremath{\mathrm{\,Me\kern -0.1em V}}\xspace}
\newcommand{\kev}{\ensuremath{\mathrm{\,ke\kern -0.1em V}}\xspace}
\newcommand{\ev}{\ensuremath{\mathrm{\,e\kern -0.1em V}}\xspace}
\newcommand{\gevc}{\ensuremath{{\mathrm{\,Ge\kern -0.1em V\!/}c}}\xspace}
\newcommand{\mevc}{\ensuremath{{\mathrm{\,Me\kern -0.1em V\!/}c}}\xspace}
\newcommand{\gevcc}{\ensuremath{{\mathrm{\,Ge\kern -0.1em V\!/}c^2}}\xspace}
\newcommand{\gevgevcccc}{\ensuremath{{\mathrm{\,Ge\kern -0.1em V^2\!/}c^4}}\xspace}
\newcommand{\mevcc}{\ensuremath{{\mathrm{\,Me\kern -0.1em V\!/}c^2}}\xspace}

\def\mum  {\ensuremath{{\,\upmu\rm m}}\xspace}

\def\invfb   {\ensuremath{\mbox{\,fb}^{-1}}\xspace}

\newcommand{\chisq}{\ensuremath{\chi^2}\xspace}

\newcommand{\chisqip}{\ensuremath{\chi^2_{\rm IP}}\xspace}
\newcommand{\chisqvs}{\ensuremath{\chi^2_{\rm VS}}\xspace}

\def\gsim{{~\raise.15em\hbox{$>$}\kern-.85em
          \lower.35em\hbox{$\sim$}~}\xspace}
\def\lsim{{~\raise.15em\hbox{$<$}\kern-.85em
          \lower.35em\hbox{$\sim$}~}\xspace}

\def\pt         {\mbox{$p_{\rm T}$}\xspace}

\def\evtgen     {\mbox{\textsc{EvtGen}}\xspace}

\def\geant      {\mbox{\textsc{Geant4}}\xspace}

\def\photos     {\mbox{\textsc{Photos}}\xspace}

\def\pythia     {\mbox{\textsc{Pythia}}\xspace}

\def\tell1  {TELL1\xspace}
\def\ukl1   {UKL1\xspace}

\def\br{{\cal{B}}}

\def\eff{\epsilon}

\def\eff{\epsilon}

\usepackage{cite} 
\usepackage{mciteplus}

\usepackage{longtable} 

\begin{document}

\renewcommand{\thefootnote}{\fnsymbol{footnote}}
\setcounter{footnote}{1}


\begin{titlepage}
\pagenumbering{roman}

\vspace*{-1.5cm}
\centerline{\large EUROPEAN ORGANIZATION FOR NUCLEAR RESEARCH (CERN)}
\vspace*{1.5cm}
\noindent
\begin{tabular*}{\linewidth}{lc@{\extracolsep{\fill}}r@{\extracolsep{0pt}}}
\ifthenelse{\boolean{pdflatex}}
{\vspace*{-2.7cm}\mbox{\!\!\!\includegraphics[width=.14\textwidth]{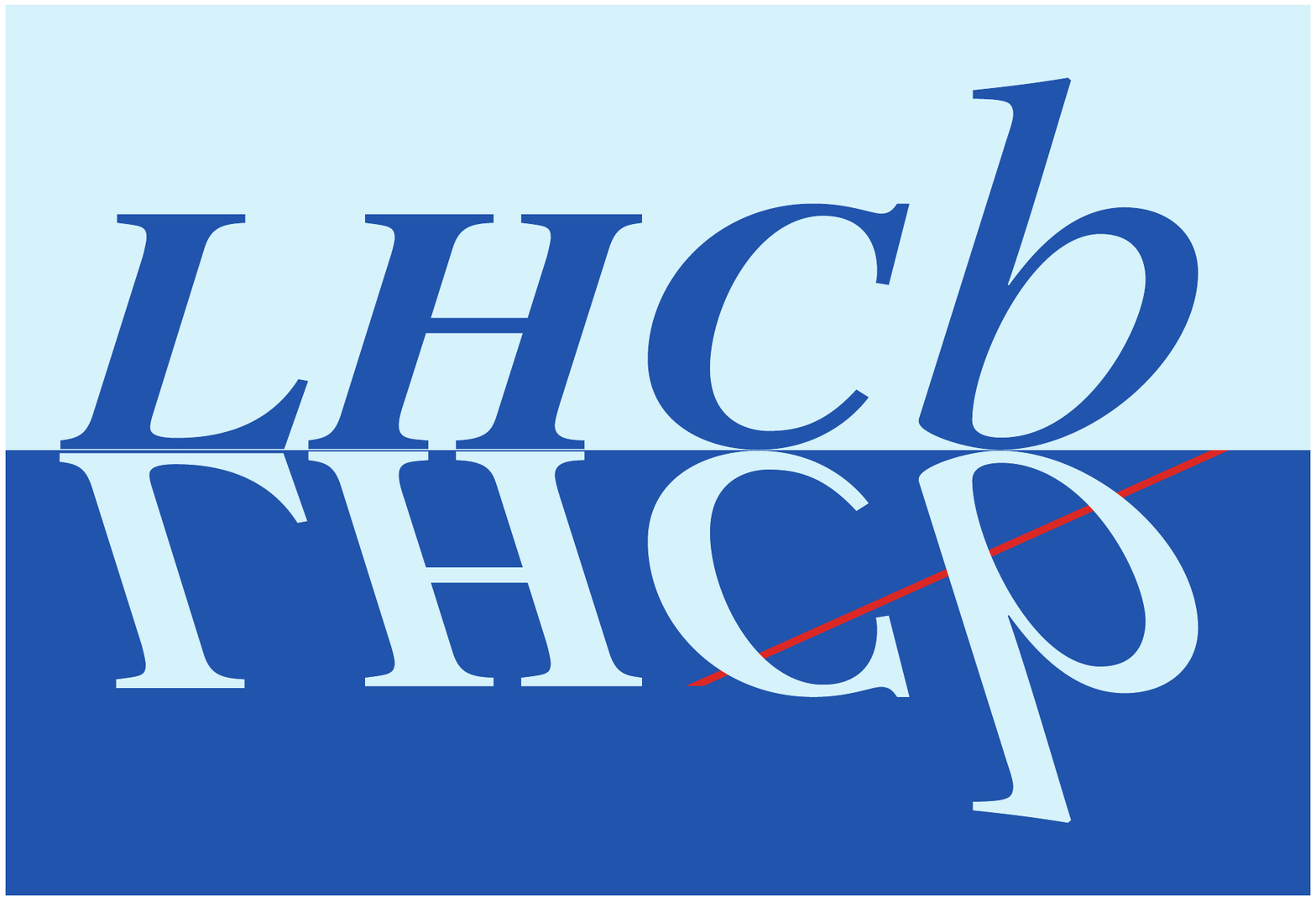}} & &}%
{\vspace*{-1.2cm}\mbox{\!\!\!\includegraphics[width=.12\textwidth]{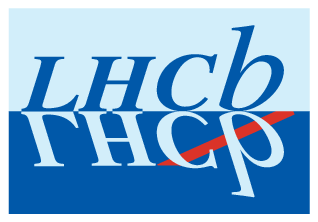}} & &}%
\\
 & & CERN-PH-EP-2015-277 \\  
 & & LHCb-PAPER-2015-047 \\  
 & & \today \\ 
 & & \\
\end{tabular*}

\vspace*{1.0cm}

{\bf\boldmath\huge
\begin{center}
  Evidence for the strangeness-changing weak decay $\Xibm\to\Lb\pim$ 
\end{center}
}

\vspace*{1.0cm}

\begin{center}
The LHCb collaboration\footnote{Authors are listed at the end of this letter.}
\end{center}

\vspace{\fill}

\begin{abstract}
  \noindent
Using a $pp$ collision data sample corresponding to an integrated luminosity of 3.0~fb$^{-1}$,
collected by the LHCb detector, we present the first search for the strangeness-changing weak decay $\Xibm\to\Lb\pim$. 
No $b$ hadron decay of this type has been seen before. A signal for this decay,
corresponding to a significance of 3.2 standard deviations, is reported.
The relative rate is measured to be
\begin{align*}
\frac{f_{\Xibm}}{f_{\Lb}}\br(\Xibm\to\Lb\pim) = (5.7\pm1.8^{+0.8}_{-0.9})\times10^{-4},
\end{align*}
\noindent where $f_{\Xibm}$ and $f_{\Lb}$ are the $b\to\Xibm$ and $b\to\Lb$ fragmentation fractions, 
and $\br(\Xibm\to\Lb\pim)$ is
the branching fraction.  Assuming $f_{\Xibm}/f_{\Lb}$ is bounded between 0.1 and 0.3, the
branching fraction $\br(\Xibm\to\Lb\pim)$ would lie in the range from $(0.57\pm0.21)\%$ to
$(0.19\pm0.07)\%$.
\end{abstract}

\vspace*{2.0cm}

\begin{center}
  Published in Phys.~Rev.~Lett.
\end{center}

\vspace{\fill}

{\footnotesize 
\centerline{\copyright~CERN on behalf of the \lhcb collaboration, licence \href{http://creativecommons.org/licenses/by/4.0/}{CC-BY-4.0}.}}
\vspace*{2mm}

\end{titlepage}


\newpage
\setcounter{page}{2}
\mbox{~}

\cleardoublepage


\renewcommand{\thefootnote}{\arabic{footnote}}
\setcounter{footnote}{0}



\pagestyle{plain} 
\setcounter{page}{1}
\pagenumbering{arabic}


%
Measurements of the lifetimes of beauty baryons provide an important test of Heavy Quark Effective Theory 
(HQET)~\cite{Khoze:1983yp,Bigi:1991ir,Bigi:1992su,Blok:1992hw,Blok:1992he,Neubert:1997gu,Uraltsev:1998bk,Bellini:1996ra}
in which it is predicted that the
decay width is dominated by the weak decay of the heavy $b$ quark. Large samples of $b$ baryons have been collected by LHCb,
enabling precise measurements of their masses and 
lifetimes~\cite{LHCb-PAPER-2014-003,LHCb-PAPER-2014-010,LHCb-PAPER-2014-021,LHCb-PAPER-2014-048}, which
are generally in good agreement with HQET predictions. Recently, it has been noted~\cite{Li:2014ada,Mannel:2015,Voloshin:2000et,Sinha:1999tc}
that for the $\Xibm$ and $\Xibz$ baryons,
the weak decay of the $s$ quark could contribute about 1\% to the total decay width. 
It has also been argued~\cite{Li:2014ada} that 
if the light diquark system has $J^P=0^+$ and exhibits the diquark correlations suggested in Refs.~\cite{Shifman:2005wa,Dosch:1988hu}, 
this could enhance the contribution from the weak decay of the $s$ quark in the $\Xibm$ ($\Xibz$) baryon to
a level that ranges from 2\% to 8\% (1\% to 4\%). 
Such a large rate would affect the comparison between HQET 
predictions and measurements of the $\Xibm$ and $\Xibz$ lifetimes. 

These ideas can be tested by studying the decay $\Xibm\to\Lb\pim$, in which the $s$ quark in the $\Xibm~(bds)$ 
undergoes a $s\to u\bar{u} d$ weak transition to a $\Lb$ ($bud$) baryon and a $\pim$ meson.
A measurement of the rate of this process would provide valuable experimental input
on the size of the aforementioned contributions to the $\Xibm$ decay width, as well as on the 
$J^P=0^+$ diquark potential.

We present a search for the decay $\Xibm\to\Lb\pim$, where the $\Lb$ baryon is reconstructed through
its decay to $\Lc\pim$, with $\Lc\to p\Km\pip$. The signal yield is normalized
with respect to the total number of $\Lb$ decays reconstructed in the same final state.
Charge conjugate processes are implied throughout. The quantity that
is measured is
\begin{align}
r_s\equiv\frac{f_{\Xibm}}{f_{\Lb}}\br(\Xibm\to\Lb\pim) = \frac{N(\Xibm\to\Lb\pim)}{N(\Lb)}\eff_{\rm rel}
\end{align}
\noindent where $f_{\Xibm}$ and $f_{\Lb}$ are the $b\to\Xibm$ and $b\to\Lb$ fragmentation fractions,
$N(\Xibm\to\Lb\pim)$ and $N(\Lb)$ are the signal yields and $\eff_{\rm rel}$ is the relative efficiency
between the normalization and signal modes. 
The signal for the $\Xibm\to\Lb\pim$ decay is a narrow peak at $38.8\pm0.5$\mevcc~\cite{LHCb-PAPER-2014-048}
in the spectrum of the mass difference, $\delta m\equiv M(\Lb\pim)-M(\Lb)-m_{\pi}$,
where $M(\Lb\pim)$ and $M(\Lb)$ are the invariant masses of the respective candidates, and $m_{\pi}$ is the
$\pim$ mass~\cite{PDG2014}.

The measurement uses proton-proton ($pp$) collision data samples
collected by the LHCb experiment, corresponding to an integrated luminosity of 3.0\invfb, of which 1.0\invfb 
was recorded at a center-of-mass energy of 7\tev and 2.0\invfb at 8\tev. 

The \lhcb detector~\cite{Alves:2008zz} is a single-arm forward
spectrometer covering the \mbox{pseudorapidity} range $2<\eta <5$,
designed for the study of particles containing \bquark or \cquark
quarks. The detector includes a high-precision tracking system,
which provides a momentum measurement with relative uncertainty of about 0.5\% from
2$-$100~\gevc and an impact parameter resolution of 20\mum for
particles with large transverse momentum (\pt). The polarity of the dipole magnet is 
reversed periodically throughout data-taking to reduce asymmetries in the detection of charged particles.
Ring-imaging Cherenkov detectors~\cite{LHCb-DP-2012-003}
are used to distinguish different types of charged hadrons. Photon, electron and
hadron candidates are identified using a calorimeter system, which is followed by
detectors to identify muons~\cite{LHCb-DP-2012-002}. 

The trigger~\cite{LHCb-DP-2012-004} consists of a
hardware stage, based on information from the calorimeter and muon
systems, and a software stage, which applies a full event
reconstruction~\cite{LHCb-DP-2012-004,BBDT}. 
The software trigger requires a two-, three- or four-track
secondary vertex that is significantly displaced from the primary $pp$
interaction vertices~(PVs) and whose tracks have a large scalar \pt sum.
At least one track should have $\pt>1.7\gevc$ and 
be inconsistent with coming from any of the PVs. The signal candidates 
are also required to pass a multivariate software trigger
selection algorithm~\cite{BBDT}. 

Proton-proton collisions are simulated using
\pythia~\cite{Sjostrand:2007gs} with a specific \lhcb
configuration~\cite{LHCb-PROC-2010-056}.  Decays of hadronic particles
are described by \evtgen~\cite{Lange:2001uf}, in which final-state
radiation is generated using \photos~\cite{Golonka:2005pn}. The
interaction of the generated particles with the detector, and its
response, are implemented using the \geant toolkit~\cite{Allison:2006ve, *Agostinelli:2002hh} as described in
Ref.~\cite{LHCb-PROC-2011-006}. 

Candidate $\Lb$ decays are formed by combining 
$\Lc\to p\Km\pip$ and $\pim$ candidates in a kinematic fit~\cite{Hulsbergen:2005pu}. 
The selection criteria are identical to those used in Ref.~\cite{LHCb-PAPER-2014-048}, except that no requirement
is made on the particle identification (PID) information for the $\pim$ candidate. 
For each combination of a $\Lb$ candidate and a PV in the event, the quantity $\chisqip$ is computed, defined to be the difference in
$\chisq$ of the PV fit when the $\Lb$ particle is included or excluded from the fit. The $\Lb$ candidate
is assigned to the PV with the smallest $\chisqip$.

Right-sign (RS) $\Xibm\to\Lb\pim$ candidates are obtained by combining a $\Lb$ 
candidate with mass in the range 5560--5680\mevcc with a $\pim$ candidate, and
wrong-sign (WS) candidates are likewise formed from $\Lb\pip$ combinations.
The pions are required to have $\pt>100$\mevc, and to have PID information 
consistent with a $\pipm$ meson. Because these pions are generally consistent with emanating from the
PV, the PID requirement helps to suppress background from other particle types.
A second kinematic fit is used to compute $\delta m$; it exploits both vertex and invariant 
mass constraints, requiring for the latter that the invariant masses of the $p\Km\pip$ and $\Lc\pim$ 
systems are equal to the known $\Lc$ and $\Lb$ masses.

Three boosted decision tree (BDT) multivariate discriminants~\cite{Breiman, AdaBoost} are used to suppress background,
one for the normalization mode (BDT1), and two for the signal mode (BDT2 and BDT3).
BDT1 is used specifically to suppress the combinatorial background contribution in
the $\Lb$ normalization mode. Five input variables
are used: the $\chi^2$ of the $\Lb$ kinematic fit; the $\chisqip$ of the $\Lb$, $\Lc$ and $\pim$ candidates; and
the $\chisqvs$ of the $\Lb$ candidate. Here, $\chisqvs$ is the difference between the $\chisq$ of the PV fit 
with and without the $\Lb$ daughter particles included in the fit. 
A large $\chisqvs$ indicates that the $\Lb$ decay vertex is
well separated from its associated PV. Simulated $\Lb\to\Lc\pim$ decays are used
to model the signal distributions of the BDT1 input variables, and candidates with $M(\Lc\pim)>5700\mevcc$  
are used to model the corresponding background spectra.
A loose selection on the BDT1 output is applied, which provides an efficiency of $(98.6\pm0.5)\%$, while
reducing the background by a factor of four. 
\begin{figure}[tb]
\centering
\includegraphics[width=0.68\textwidth]{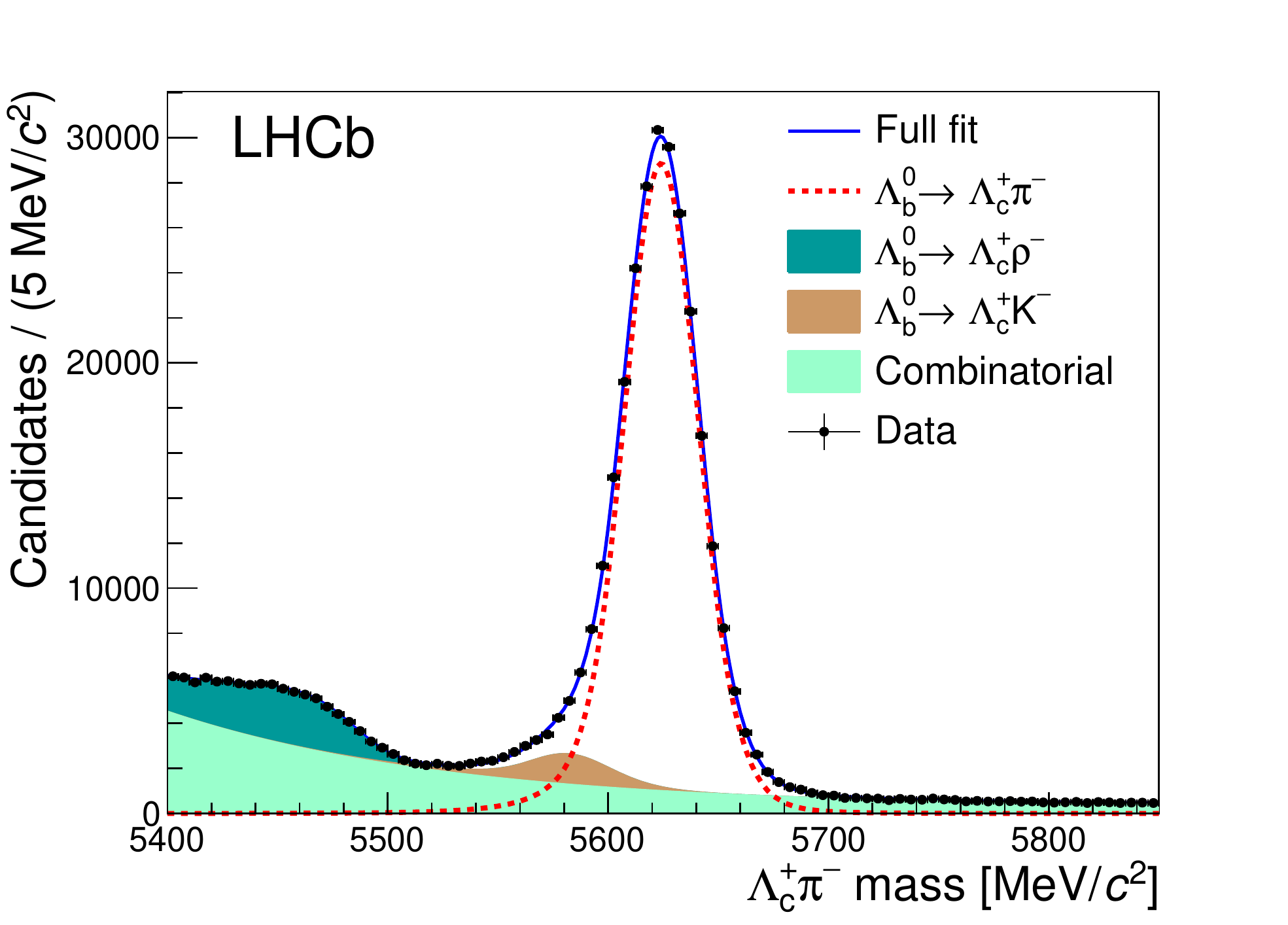}
\caption{\small{Invariant mass spectrum for selected $\Lb\to\Lc\pim$ candidates in data.}}
\label{fig:Lb2LcPi}
\end{figure}

The invariant mass spectrum of selected $\Lc\pim$ candidates is displayed in Fig.~\ref{fig:Lb2LcPi}.
The yield is determined from an unbinned
extended maximum likelihood fit using the signal and background shapes as described in Ref.~\cite{LHCb-PAPER-2014-021}.
The fitted number of $\Lb\to\Lc\pim$ decays is $(256.7\pm0.6)\times10^3$, and the fraction 
$N(\Lb\to\Lc\Km)/N(\Lb\to\Lc\pim)=(5.9\pm0.2)$\%, where the uncertainties are statistical only.
In the mass region 5560--5680\mevcc, the 
fitted yields of $\Lb\to\Lc\pim$ and $\Lb\to\Lc\Km$ decays are 253\,300 and 11\,700, respectively. Since misidentified 
$\Lb\to\Lc\Km$ signal decays also contribute to the $\Xibm\to\Lb\pim$ signal, they are included
in the total normalization mode yield. Thus the signal yield for the normalization mode
is $(265\pm1)\times10^{3}$.

The second BDT (BDT2) has the same purpose as BDT1, except that it is applied to the $\Lb$ candidates within the
$\Xibm\to\Lb\pim$ sample. This alternate BDT is needed since the lifetime of the $\Xibm$ baryon is about the
same as that of the $\Lb$ baryon, thus leading to larger typical values of $\chisqvs$ compared to the inclusively produced $\Lb$ sample.
A similar training to that of BDT1 is performed, except that the signal distributions are taken from 
simulated $\Xibm\to\Lb\pim$ decays. A loose selection on the BDT2 output yields an efficiency of $(99.0\pm0.5)\%$. 

The third BDT (BDT3) is used to distinguish real $\Xibm\to\Lb\pim$ decays from $\Lb$ baryons combined with a random $\pim$
candidate. Because of the small energy release in the $\Xibm\to\Lb\pim$ decay, the $\Lb$ and $\pim$ directions are nearly collinear with that 
of the $\Xibm$. This makes it difficult to identify the $\Lb$ and $\pim$ daughters as particles produced at a secondary vertex.
The input variables used in BDT3 are the flight distance and $\chisqvs$ of the $\Xibm$ candidate, the $\chisqvs$ of
the $\Lb$ candidate, and the $\chisqip$ and $\pt$ of the low momentum (slow) $\pim$ daughter of the $\Xibm$ candidate. 
The signal distributions of these variables are taken from simulated $\Xibm\to\Lb\pim$ decays, and the background spectra are 
taken from WS candidates that have $34<\delta m<44$\mevcc. Separate training and test
samples were compared and showed no bias due to overtraining.

A loose selection on the BDT3 output is applied, rejecting about 3\% of the expected signal events. 
The selected events are divided into two signal-to-background (S/B) regions according to the BDT3 output: 
a high S/B region and a low S/B region. The split between the high and 
low S/B regions is chosen to provide optimal expected sensitivity. 
The expected ratio of yields in the low S/B to high S/B regions is 1.60, which is fixed in the fit to data. 

An event may have more than one $\Xibm$ candidate, almost always due to a single $\Lb$ candidate being combined with
more than one $\pim$ candidate.  The average number of candidates in events that contain a candidate in the low S/B region is 1.35, 
and 1.02 in events that contain a candidate in the high S/B region. All candidates are kept. Potential bias on the signal 
yield determination due to this choice was investigated, and none was found.

Four disjoint subsamples of data are used in the fits, split by charge (RS, WS) and by S/B region (low, high).
Including the WS data allows additional constraints on the shape of the combinatorial background, and also 
provides a consistency check that the signal yield in the $\Lb\pip$ mode is consistent with zero.
In these four $\delta m$ spectra we allow for three contributions: a $\Xibm\to\Lb\pim$ signal, strong decays of
$\Sigmares_b^{(*)\pm}\to\Lb\pi^{\pm}$ resonances, and combinatorial background. The low S/B region contains
almost all of the $\Sigmares_b^{(*)\pm}\to\Lb\pi^{\pm}$ signal decays. The primary reason for including the low
S/B regions is that they contain almost all of the $\Sigmares_b^{(*)\pm}\to\Lb\pi^{\pm}$ signal decays. This leads to
tighter constraints on the $\Sigmares_b^{(*)\pm}\to\Lb\pi^{\pm}$ mass shapes in the high S/B region, since the
shape parameters are common to the low and high S/B regions. 
A simultaneous unbinned extended maximum likelihood fit is performed to the four $\delta m$ spectra, in the range
2--122\mevcc, using the signal and background shapes discussed below.

The $\delta m$ signal shape is obtained from simulated $\Xibm\to\Lb\pim$ decays, allowing for different
signal shapes in the low and high S/B regions. 
Each sample is fit to the sum of two Gaussian functions with a common mean value.
The shapes are slightly different, but the average resolution, given as the weighted average of the two Gaussian widths, 
is 1.57\mevcc in both cases. All signal shape parameters are fixed in fits to data, including the mean, which is
fixed to $M(\Xibm)-M(\Lb)-m_{\pim}=38.8\mevcc$~\cite{LHCb-PAPER-2014-048}.
A scale factor of 1.10 is applied to the widths to account for slightly worse resolution in data than simulation,
as determined from a study of the $\delta m$ resolution in $\Dstarp\to\Dz\pip$ decays~\cite{LHCb-CONF-2013-003}. 
Variations in this value are considered as a source of systematic uncertainty.

The contributions from the $\Sigmares_b^{\pm}$ and $\Sigmares_b^{*\pm}$ resonances are each modeled using a relativistic Breit Wigner 
shape~\cite{Jackson:1964zd}.
Each of them is convolved with a resolution function obtained from simulated $\Sigmares_b^{(*)-}$ decays,
and is parameterized as the sum of three Gaussian distributions with a common mean. 
The average resolution is 1.97\mevcc for $\Sigmares_b^-$ and
2.25\mevcc for $\Sigmares_b^{*-}$.
The $\Sigmares^{(*)\pm}$ masses and natural widths are freely varied in the fit to data, but the Gaussian widths
are fixed and include a scale factor of 1.10, as indicated previously. The masses and widths of the
$\Sigmares_b^{(*)\pm}$ resonances are being studied in a separate analysis.

The combinatorial background is described by the threshold function
\begin{align}
f_{\rm back}(\delta m) \propto (\delta m)^A(1-e^{-\delta m/C}),
\end{align}
\noindent where the parameters $A$ and $C$ are freely varied in the fit to data. One set of parameters is used for the
low S/B region, and a separate set for the high S/B region. For each S/B region, the RS and WS spectra share a common set of parameters.

\label{sec:fitdata}

\begin{figure}[tb]
\centering
\includegraphics[width=0.48\textwidth]{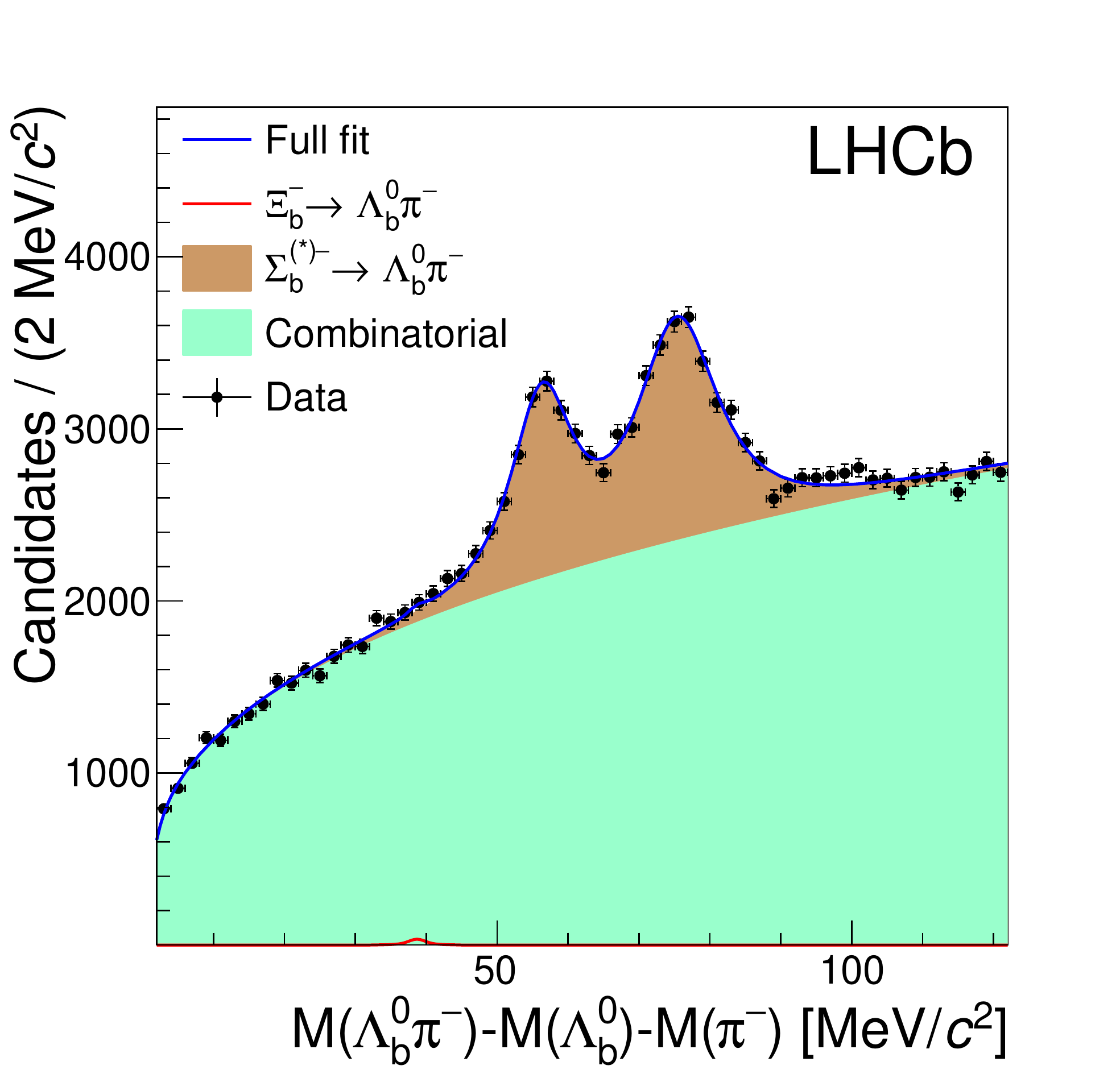}
\includegraphics[width=0.48\textwidth]{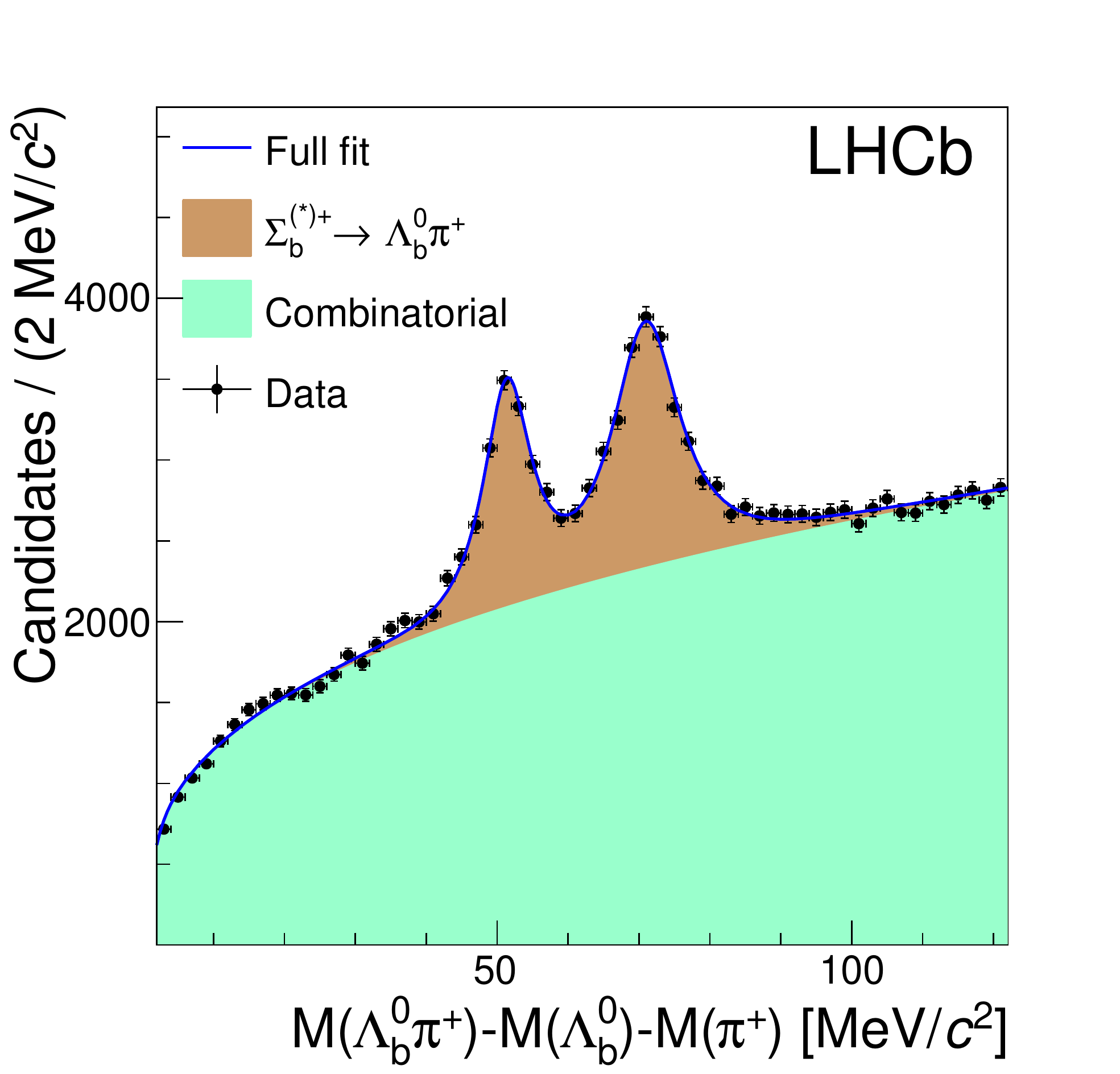}
\includegraphics[width=0.48\textwidth]{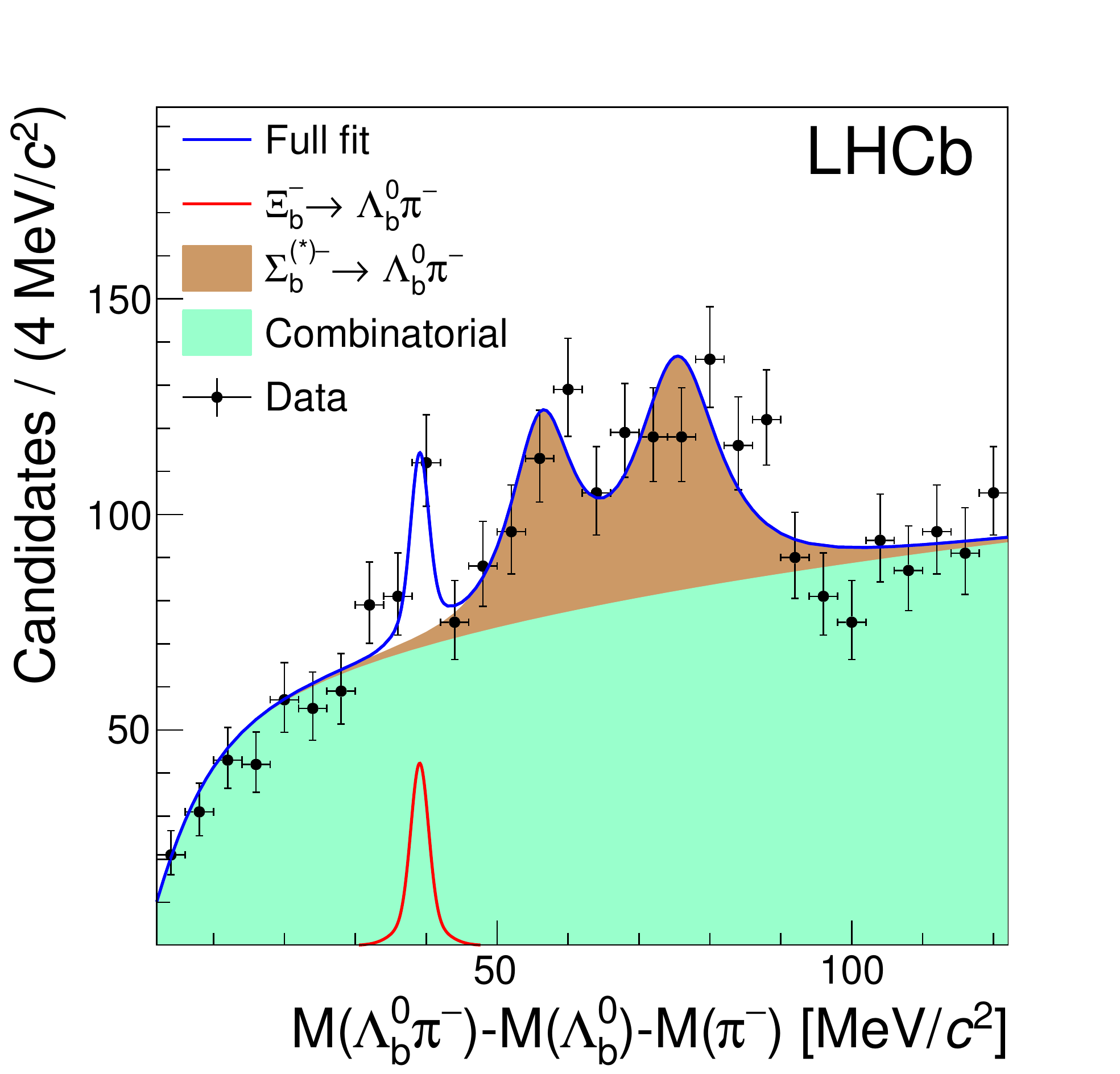}
\includegraphics[width=0.48\textwidth]{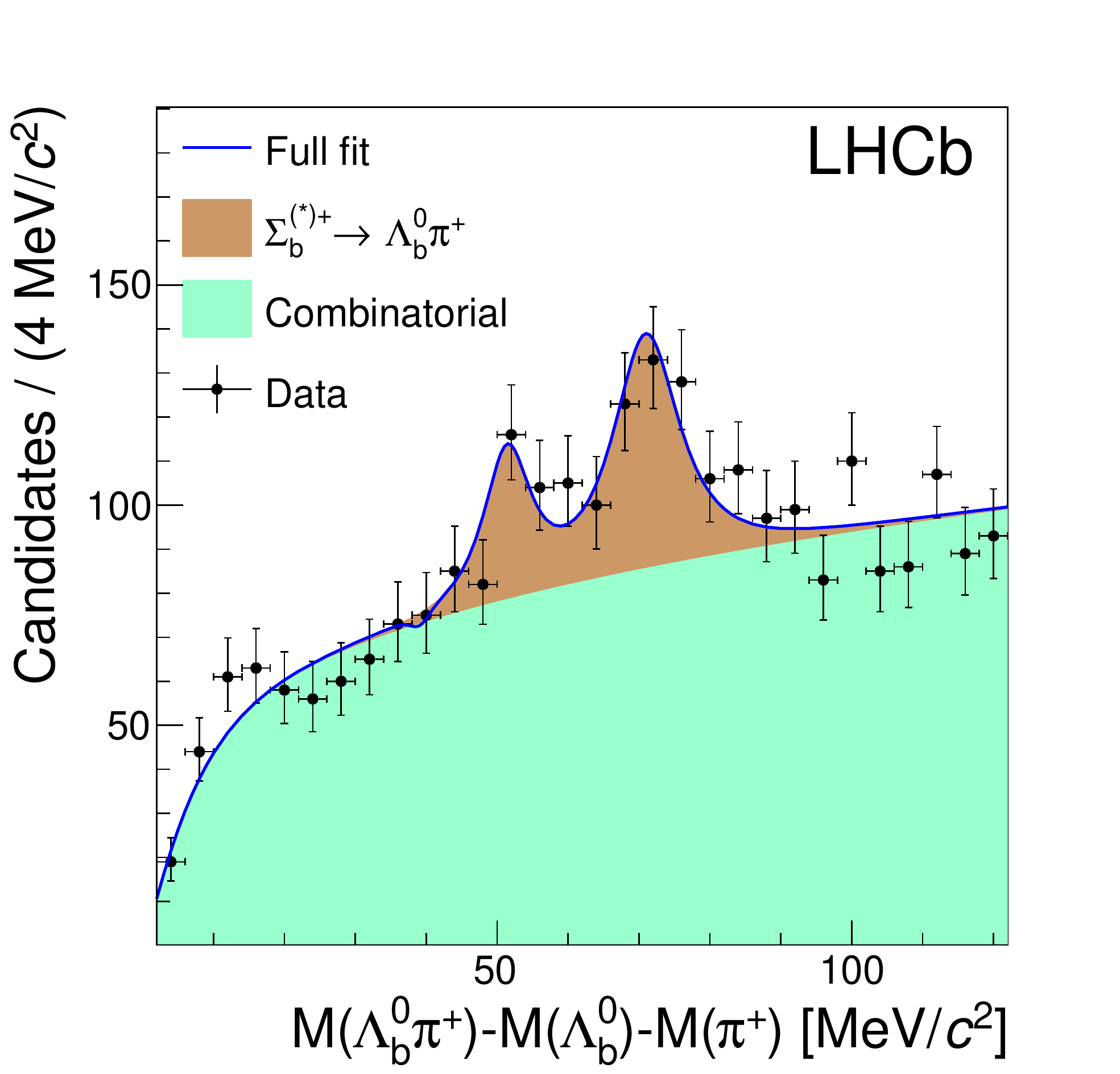}
\caption{\small{Fit to the $\delta m$ spectra in data: (top left) RS low S/B, (top right) WS low S/B,
(bottom left) RS high S/B and (bottom right) WS high S/B.}}
\label{fig:Xib2LbPi}
\end{figure}

The resulting mass fits are shown in Fig.~\ref{fig:Xib2LbPi}. 
\noindent The $\Sigmares_b^{\pm}$ and $\Sigmares_b^{*\pm}$ signals appear prominently, and are constrained by the 
data in the low S/B spectra (top pair of plots).
The data show an enhancement at the expected $\delta m$ value for the $\Xibm\to\Lb\pim$ decay
in the RS high S/B region, but no such excess is seen in the corresponding WS sample.
The total fitted signal yields for the RS and WS samples are $103\pm33$ and $-7\pm28$, respectively.

The relative efficiency between the normalization and signal modes can be expressed as
\begin{align}
\eff_{\rm rel}\equiv\frac{\eff_{\Lb}}{\eff_{\Xibm}}=\frac{\eff_{\rm rel}^{\rm acc}\cdot\eff_{\rm rel}^{\rm rec}\cdot\eff_{\rm rel}^{\rm BDT1,2}}{\eff_{\Xibm\rm -only}}
\end{align}
\noindent where $\eff_{\rm rel}^{\rm acc}=1.03\pm0.01$ is the relative efficiency for all of the stable daughter particles 
to be within the LHCb acceptance; $\eff_{\rm rel}^{\rm rec}=1.38\pm0.02$ is the relative efficiency for reconstruction and
selection, including the $\pt>100$\mevc requirement on the $\pim$ meson; $\eff_{\rm rel}^{\rm BDT1,2}=1.00\pm0.01$ is the
relative efficiency of the BDT1 and BDT2 selections; and $\eff_{\Xibm\rm -only}=0.95\pm0.01$ includes the BDT3 requirement
and the PID selection criteria on the $\pim$ candidate. The relative efficiencies are obtained from simulated
$\Xibm\to\Lb\pim$ events and inclusively produced $\Lb\to\Lc\pim$ decays, except for the PID requirements, which are taken from
$\Dstarp\to\Dz\pip$ calibration data. The relative efficiency is therefore $1.47\pm0.03$. 

Several sources of systematic uncertainty affect the signal yield determination and thus the
signal significance. Additional sources of systematic uncertainty contribute to the
determination of $r_s$. The uncertainties are summarized in Table~\ref{tab:syst}.
In the default fit, the $\Xibm$ signal peak position is fixed to the nominal value of $\delta m=38.8$\mevcc,
which has an uncertainty of 0.5\mevcc. We therefore refit the data with the peak position shifted by $\pm$0.5\mevcc, obtaining 
changes of $-6.4\%$ and $+4.9\%$ in the yield. These values are assigned as a systematic error.
Uncertainty in the signal yield due to the fixed mass resolution scale factor of 1.10 is investigated by
varying it by $\pm$0.05, and we assign
the average change in yield of 3.0\% as a systematic error. Variations in the corresponding
scale factor for the $\Sigmares_b^{(*)-}$ resonances were investigated, and were found to have 
negligible impact on the $\Xibm\to\Lb\pim$ signal yield. Different choices for the fit range and
the combinatorial background function were investigated, and among these fit variations, a
maximum shift in the signal yield of 12.6\% was found. The full difference is assigned as a 
systematic uncertainty.

Additional systematic uncertainties that affect $r_s$ include the relative efficiency
between the low and high S/B regions, the slow $\pim$ detection efficiency,
and the yield of $\Lb$ decays.
In comparing the BDT1 distributions for $\Lb\to\Lc\pim$ signal in data and simulation,
as well as the background distributions for BDT3 in data and simulation,
the relative efficiencies do not vary by more than 2\% for any BDT selection.
We therefore assign 2\% as a systematic uncertainty.
The $\pim$ meson from the $\Xibm$ decay must be reconstructed and have $\pt>100\mevc$. 
The tracking efficiency uncertainty is assessed using data-driven techniques~\cite{LHCb-DP-2013-002}, and is less
than 1.6\%.  The uncertainty due to the $\pt$ requirement
is estimated by interpolating the \pt spectrum from 100\mevc to zero in simulated decays, 
and assuming that the fraction of signal events in this \pt region in 
data could differ from the simulated fraction by as much as 25\%. This leads to a model uncertainty of 1.7\%. 
Thus, an uncertainty of 2.3\% is assigned to the detection of the $\pim$ from the $\Xibm$ decay.

For the number of $\Lb$ signal events, we assign a 1.0\% uncertainty, which includes both the
statistical component and a systematic uncertainty due to the signal and background shapes used to
fit the $\Lc\pim$ mass spectrum. 

\begin{table*}[tb]
\begin{center}
\caption{\small{Relative systematic uncertainties (in percent) on the signal yield determination and on the quantity 
$\left(f_{\Xibm}/f_{\Lb}\right)\br(\Xibm\to\Lb\pim)$, as described in the text.}}
\begin{tabular}{lc}
\hline\hline
Source  & Value (\%) \\
\hline
\\ [-2.0ex]
Mean $\delta m$     & $^{+4.9}_{-6.4}$ \\
\\ [-2.2ex]
Signal resolution   & 3.0 \\
Combinatorial background shape    & 12.6 \\
\hline
$\eff({\rm High~S/B})/\eff({\rm Low~S/B})$       &  2.0  \\
Slow $\pim$ efficiency            &  2.3 \\
$\Lb$ normalization mode yield    &  1.0 \\
Simulated sample size             &  2.1 \\
\hline
\\ [-1.8ex]
Total for signal significance            & $^{+13.9}_{-14.5}$ \\
\\ [-1.8ex]
Total for $r_s$            & $^{+14.4}_{-15.0}$ \\
\\ [-2.0ex]
\hline\hline
\end{tabular}
\label{tab:syst}
\end{center}
\end{table*}

To check the robustness of the signal, the data were partitioned into different subsamples and the fitted
yields in each were determined independently. The subsamples consisted of only 2012 data ($\sim$2/3 of the data set), 
only negative magnet polarity data ($\sim$50\% of the data sample), and only $\Lb\pim$ data, not $\Lbbar\pip$  
(expect $\sim$50\%).
In all three cases the signal yields are compatible with expectations. Other robustness checks were also performed,
such as placing a stringent PID requirement on the $\pim$, fitting only the RS data, and using only raw invariant masses 
(without the full kinematic fit). Upward and downward variations are observed, but in all cases, the fitted yields 
are consistent with expectations.

The significance of the signal is computed with Wilks's theorem~\cite{Wilks:1938dza}.
The systematic uncertainty is included by convolving the likelihood function with a bifurcated Gaussian distribution
whose widths are given by the asymmetric uncertainties in Table~\ref{tab:syst}, which leads to a significance of 3.2$\sigma$.
We thus have evidence for the $\Xibm\to\Lb\pim$ decay.

With the yields and relative efficiencies presented previously, we find
\begin{align*}
\frac{f_{\Xibm}}{f_{\Lb}}\br(\Xibm\to\Lb\pim) = (5.7\pm1.8\,^{+0.8}_{-0.9})\times10^{-4},
\end{align*}
where the uncertainties are statistical and systematic, respectively.
To assess what this value implies in terms of $\br(\Xibm\to\Lb\pim)$, we consider a plausible range for
$f_{\Xibm}/f_{\Lb}$ from 0.1--0.3, based on measured production rates of other strange 
particles relative to their non-strange counterparts~\cite{LHCb-PAPER-2011-037,LHCb-PAPER-2012-041,LHCb-PAPER-2012-037,LHCb-PAPER-2015-041,PDG2014}. 
Assuming $f_{\Xibm}/f_{\Lb}$ is bounded between 0.1 and 0.3, the
branching fraction $\br(\Xibm\to\Lb\pim)$ would be in the range from $(0.57\pm0.21)\%$ to
$(0.19\pm0.07)\%$. 

In summary, we present the first evidence for the $\Xibm\to\Lb\pim$ decay, which is mediated by the
weak transition of the $s$ quark. With the above assumptions for $f_{\Xibm}/f_{\Lb}$,
the measured value for $\br(\Xibm\to\Lb\pim)$ is consistent with the range of 0.19--0.76\%, 
predicted in Ref.~\cite{Mannel:2015} assuming the diquark transitions have roughly the same
weak amplitude as in $B$, $D$, and $K$ meson decays. The results are also consistent with the
value of 0.57--0.62\%, obtained using either a current algebra or pole model approach,
but are inconsistent with the values of 0.01\% and 0.012\% using the factorization approximation or
the quark line approach~\cite{Sinha:1999tc}.
The measured value of $\br(\Xibm\to\Lb\pim)$ disfavors a large enhancement to the decay rate of 
$\Xibm$ baryons from the $s\to u\bar{u}d$ transition, which could occur if the
short-distance correlations within the $J^P=0^+$ diquark system are enhanced,
as suggested in Refs.~\cite{Li:2014ada,Shifman:2005wa,Dosch:1988hu}.


\section*{Acknowledgements}

\noindent We express our gratitude to our colleagues in the CERN
accelerator departments for the excellent performance of the LHC. We
thank the technical and administrative staff at the LHCb
institutes. We acknowledge support from CERN and from the national
agencies: CAPES, CNPq, FAPERJ and FINEP (Brazil); NSFC (China);
CNRS/IN2P3 (France); BMBF, DFG and MPG (Germany); INFN (Italy);
FOM and NWO (The Netherlands); MNiSW and NCN (Poland); MEN/IFA (Romania);
MinES and FANO (Russia); MinECo (Spain); SNSF and SER (Switzerland);
NASU (Ukraine); STFC (United Kingdom); NSF (USA).
We acknowledge the computing resources that are provided by CERN, IN2P3 (France), KIT and DESY (Germany), INFN (Italy), SURF (The Netherlands), PIC (Spain), GridPP (United Kingdom), RRCKI (Russia), CSCS (Switzerland), IFIN-HH (Romania), CBPF (Brazil), PL-GRID (Poland) and OSC (USA). We are indebted to the communities behind the multiple open
source software packages on which we depend. We are also thankful for the
computing resources and the access to software R\&D tools provided by Yandex LLC (Russia).
Individual groups or members have received support from AvH Foundation (Germany),
EPLANET, Marie Sk\l{}odowska-Curie Actions and ERC (European Union),
Conseil G\'{e}n\'{e}ral de Haute-Savoie, Labex ENIGMASS and OCEVU,
R\'{e}gion Auvergne (France), RFBR (Russia), GVA, XuntaGal and GENCAT (Spain), The Royal Society
and Royal Commission for the Exhibition of 1851 (United Kingdom).

\clearpage

\begin{mcitethebibliography}{10}
\mciteSetBstSublistMode{n}
\mciteSetBstMaxWidthForm{subitem}{\alph{mcitesubitemcount})}
\mciteSetBstSublistLabelBeginEnd{\mcitemaxwidthsubitemform\space}
{\relax}{\relax}

\bibitem{Khoze:1983yp}
V.~A. Khoze and M.~A. Shifman,
  \ifthenelse{\boolean{articletitles}}{\emph{{Heavy quarks}},
  }{}\href{http://dx.doi.org/10.1070/PU1983v026n05ABEH004398}{Sov.\ Phys.\
  Usp.\  \textbf{26} (1983) 387}\relax
\mciteBstWouldAddEndPuncttrue
\mciteSetBstMidEndSepPunct{\mcitedefaultmidpunct}
{\mcitedefaultendpunct}{\mcitedefaultseppunct}\relax
\EndOfBibitem
\bibitem{Bigi:1991ir}
I.~I. Bigi and N.~G. Uraltsev,
  \ifthenelse{\boolean{articletitles}}{\emph{{Gluonic enhancements in
  non-spectator beauty decays - an inclusive mirage though an exclusive
  possibility}},
  }{}\href{http://dx.doi.org/10.1016/0370-2693(92)90066-D}{Phys.\ Lett.\
  \textbf{B280} (1992) 271}\relax
\mciteBstWouldAddEndPuncttrue
\mciteSetBstMidEndSepPunct{\mcitedefaultmidpunct}
{\mcitedefaultendpunct}{\mcitedefaultseppunct}\relax
\EndOfBibitem
\bibitem{Bigi:1992su}
I.~I. Bigi, N.~G. Uraltsev, and A.~I. Vainshtein,
  \ifthenelse{\boolean{articletitles}}{\emph{{Nonperturbative corrections to
  inclusive beauty and charm decays: QCD versus phenomenological models}},
  }{}\href{http://dx.doi.org/10.1016/0370-2693(92)90908-M}{Phys.\ Lett.\
  \textbf{B293} (1992) 430}, Erratum
  \href{http://dx.doi.org/10.1016/0370-2693(92)91287-J}{ibid.\   \textbf{B297}
  (1992) 477}, \href{http://arxiv.org/abs/hep-ph/9207214}{{\tt
  arXiv:hep-ph/9207214}}\relax
\mciteBstWouldAddEndPuncttrue
\mciteSetBstMidEndSepPunct{\mcitedefaultmidpunct}
{\mcitedefaultendpunct}{\mcitedefaultseppunct}\relax
\EndOfBibitem
\bibitem{Blok:1992hw}
B.~Blok and M.~Shifman, \ifthenelse{\boolean{articletitles}}{\emph{{The rule of
  discarding 1/$N_c$ in inclusive weak decays (I)}},
  }{}\href{http://dx.doi.org/10.1016/0550-3213(93)90504-I}{Nucl.\ Phys.\
  \textbf{B399} (1993) 441}, \href{http://arxiv.org/abs/hep-ph/9207236}{{\tt
  arXiv:hep-ph/9207236}}\relax
\mciteBstWouldAddEndPuncttrue
\mciteSetBstMidEndSepPunct{\mcitedefaultmidpunct}
{\mcitedefaultendpunct}{\mcitedefaultseppunct}\relax
\EndOfBibitem
\bibitem{Blok:1992he}
B.~Blok and M.~Shifman, \ifthenelse{\boolean{articletitles}}{\emph{{The rule of
  discarding 1/$N_c$ in inclusive weak decays (II)}},
  }{}\href{http://dx.doi.org/10.1016/0550-3213(93)90505-J}{Nucl.\ Phys.\
  \textbf{B399} (1993) 459}, \href{http://arxiv.org/abs/hep-ph/9209289}{{\tt
  arXiv:hep-ph/9209289}}\relax
\mciteBstWouldAddEndPuncttrue
\mciteSetBstMidEndSepPunct{\mcitedefaultmidpunct}
{\mcitedefaultendpunct}{\mcitedefaultseppunct}\relax
\EndOfBibitem
\bibitem{Neubert:1997gu}
M.~Neubert, \ifthenelse{\boolean{articletitles}}{\emph{{B decays and the
  heavy-quark expansion}},
  }{}\href{http://dx.doi.org/10.1142/9789812812667_0003}{Adv.\ Ser.\ Direct.\
  High Energy Phys.\  \textbf{15} (1998) 239},
  \href{http://arxiv.org/abs/hep-ph/9702375}{{\tt arXiv:hep-ph/9702375}}\relax
\mciteBstWouldAddEndPuncttrue
\mciteSetBstMidEndSepPunct{\mcitedefaultmidpunct}
{\mcitedefaultendpunct}{\mcitedefaultseppunct}\relax
\EndOfBibitem
\bibitem{Uraltsev:1998bk}
N.~Uraltsev, \ifthenelse{\boolean{articletitles}}{\emph{{Heavy quark expansion
  in beauty and its decays}},
  }{}\href{http://arxiv.org/abs/hep-ph/9804275}{{\tt arXiv:hep-ph/9804275}},
  also published in proceedings, Heavy Flavour Physics: A Probe of Nature's
  Grand Design, Proc. Intern. School of Physics ``Enrico Fermi'', Course
  CXXXVII, Varenna, July 7-18 1997\relax
\mciteBstWouldAddEndPuncttrue
\mciteSetBstMidEndSepPunct{\mcitedefaultmidpunct}
{\mcitedefaultendpunct}{\mcitedefaultseppunct}\relax
\EndOfBibitem
\bibitem{Bellini:1996ra}
G.~Bellini, I.~I.~Y. Bigi, and P.~J. Dornan,
  \ifthenelse{\boolean{articletitles}}{\emph{{Lifetimes of charm and beauty
  hadrons}}, }{}\href{http://dx.doi.org/10.1016/S0370-1573(97)00005-7}{Phys.\
  Rept.\  \textbf{289} (1997) 1}\relax
\mciteBstWouldAddEndPuncttrue
\mciteSetBstMidEndSepPunct{\mcitedefaultmidpunct}
{\mcitedefaultendpunct}{\mcitedefaultseppunct}\relax
\EndOfBibitem
\bibitem{LHCb-PAPER-2014-003}
LHCb collaboration, R.~Aaij {\em et~al.},
  \ifthenelse{\boolean{articletitles}}{\emph{{Precision measurement of the
  ratio of the $\Lambda_b^0$ to $\overline{B}^0$ lifetimes}},
  }{}\href{http://dx.doi.org/10.1016/j.physletb.2014.05.021}{Phys.\ Lett.\
  \textbf{B734} (2014) 122}, \href{http://arxiv.org/abs/1402.6242}{{\tt
  arXiv:1402.6242}}\relax
\mciteBstWouldAddEndPuncttrue
\mciteSetBstMidEndSepPunct{\mcitedefaultmidpunct}
{\mcitedefaultendpunct}{\mcitedefaultseppunct}\relax
\EndOfBibitem
\bibitem{LHCb-PAPER-2014-010}
LHCb collaboration, R.~Aaij {\em et~al.},
  \ifthenelse{\boolean{articletitles}}{\emph{{Measurement of the $\Xi_b^-$ and
  $\Omega_b^-$ baryon lifetimes}},
  }{}\href{http://dx.doi.org/10.1016/j.physletb.2014.06.064}{Phys.\ Lett.\
  \textbf{B736} (2014) 154}, \href{http://arxiv.org/abs/1405.1543}{{\tt
  arXiv:1405.1543}}\relax
\mciteBstWouldAddEndPuncttrue
\mciteSetBstMidEndSepPunct{\mcitedefaultmidpunct}
{\mcitedefaultendpunct}{\mcitedefaultseppunct}\relax
\EndOfBibitem
\bibitem{LHCb-PAPER-2014-021}
LHCb collaboration, R.~Aaij {\em et~al.},
  \ifthenelse{\boolean{articletitles}}{\emph{{Precision measurement of the mass
  and lifetime of the $\Xi_b^0$ baryon}},
  }{}\href{http://dx.doi.org/10.1103/PhysRevLett.113.032001}{Phys.\ Rev.\
  Lett.\  \textbf{113} (2014) 032001},
  \href{http://arxiv.org/abs/1405.7223}{{\tt arXiv:1405.7223}}\relax
\mciteBstWouldAddEndPuncttrue
\mciteSetBstMidEndSepPunct{\mcitedefaultmidpunct}
{\mcitedefaultendpunct}{\mcitedefaultseppunct}\relax
\EndOfBibitem
\bibitem{LHCb-PAPER-2014-048}
LHCb collaboration, R.~Aaij {\em et~al.},
  \ifthenelse{\boolean{articletitles}}{\emph{{Precision measurement of the mass
  and lifetime of the $\Xi_b^-$ baryon}},
  }{}\href{http://dx.doi.org/10.1103/PhysRevLett.113.242002}{Phys.\ Rev.\
  Lett.\  \textbf{113} (2014) 242002},
  \href{http://arxiv.org/abs/1409.8568}{{\tt arXiv:1409.8568}}\relax
\mciteBstWouldAddEndPuncttrue
\mciteSetBstMidEndSepPunct{\mcitedefaultmidpunct}
{\mcitedefaultendpunct}{\mcitedefaultseppunct}\relax
\EndOfBibitem
\bibitem{Li:2014ada}
X.~Li and M.~B. Voloshin, \ifthenelse{\boolean{articletitles}}{\emph{{Decays
  $\Xi_b \to \Lambda_{b} \pi$ and diquark correlations in hyperons}},
  }{}\href{http://dx.doi.org/10.1103/PhysRevD.90.033016}{Phys.\ Rev.\
  \textbf{D90} (2014) 033016}, \href{http://arxiv.org/abs/1407.2556}{{\tt
  arXiv:1407.2556}}\relax
\mciteBstWouldAddEndPuncttrue
\mciteSetBstMidEndSepPunct{\mcitedefaultmidpunct}
{\mcitedefaultendpunct}{\mcitedefaultseppunct}\relax
\EndOfBibitem
\bibitem{Mannel:2015}
S.~Faller and T.~Mannel, \ifthenelse{\boolean{articletitles}}{\emph{{Light
  quark decays in heavy hadrons}},
  }{}\href{http://arxiv.org/abs/1503.06088}{{\tt arXiv:1503.06088}}\relax
\mciteBstWouldAddEndPuncttrue
\mciteSetBstMidEndSepPunct{\mcitedefaultmidpunct}
{\mcitedefaultendpunct}{\mcitedefaultseppunct}\relax
\EndOfBibitem
\bibitem{Voloshin:2000et}
M.~B. Voloshin, \ifthenelse{\boolean{articletitles}}{\emph{{Weak decays
  $\Xi_Q\to\Lambda_Q\pi$}},
  }{}\href{http://dx.doi.org/10.1016/S0370-2693(00)00150-7}{Phys.\ Lett.\
  \textbf{B476} (2000) 297}, \href{http://arxiv.org/abs/hep-ph/0001057}{{\tt
  arXiv:hep-ph/0001057}}\relax
\mciteBstWouldAddEndPuncttrue
\mciteSetBstMidEndSepPunct{\mcitedefaultmidpunct}
{\mcitedefaultendpunct}{\mcitedefaultseppunct}\relax
\EndOfBibitem
\bibitem{Sinha:1999tc}
S.~Sinha and M.~P. Khanna,
  \ifthenelse{\boolean{articletitles}}{\emph{{Beauty-conserving
  strangeness-changing two-body hadronic decays of beauty baryons}},
  }{}\href{http://dx.doi.org/10.1142/S0217732399000705}{Mod.\ Phys.\ Lett.\
  \textbf{A14} (1999) 651}\relax
\mciteBstWouldAddEndPuncttrue
\mciteSetBstMidEndSepPunct{\mcitedefaultmidpunct}
{\mcitedefaultendpunct}{\mcitedefaultseppunct}\relax
\EndOfBibitem
\bibitem{Shifman:2005wa}
M.~Shifman and A.~Vainshtein,
  \ifthenelse{\boolean{articletitles}}{\emph{{Remarks on diquarks, strong
  binding and a large hidden QCD scale}},
  }{}\href{http://dx.doi.org/10.1103/PhysRevD.71.074010}{Phys.\ Rev.\
  \textbf{D71} (2005) 074010}, \href{http://arxiv.org/abs/hep-ph/0501200}{{\tt
  arXiv:hep-ph/0501200}}\relax
\mciteBstWouldAddEndPuncttrue
\mciteSetBstMidEndSepPunct{\mcitedefaultmidpunct}
{\mcitedefaultendpunct}{\mcitedefaultseppunct}\relax
\EndOfBibitem
\bibitem{Dosch:1988hu}
H.~G. Dosch, M.~Jamin, and B.~Stech,
  \ifthenelse{\boolean{articletitles}}{\emph{{Diquarks, {QCD} sum rules and
  weak decays}}, }{}\href{http://dx.doi.org/10.1007/BF01565139}{Z.\ Phys.\
  \textbf{C42} (1989) 167}\relax
\mciteBstWouldAddEndPuncttrue
\mciteSetBstMidEndSepPunct{\mcitedefaultmidpunct}
{\mcitedefaultendpunct}{\mcitedefaultseppunct}\relax
\EndOfBibitem
\bibitem{PDG2014}
Particle Data Group, K.~A. Olive {\em et~al.},
  \ifthenelse{\boolean{articletitles}}{\emph{{\href{http://pdg.lbl.gov/}{Review
  of particle physics}}},
  }{}\href{http://dx.doi.org/10.1088/1674-1137/38/9/090001}{Chin.\ Phys.\
  \textbf{C38} (2014) 090001}\relax
\mciteBstWouldAddEndPuncttrue
\mciteSetBstMidEndSepPunct{\mcitedefaultmidpunct}
{\mcitedefaultendpunct}{\mcitedefaultseppunct}\relax
\EndOfBibitem
\bibitem{Alves:2008zz}
LHCb collaboration, A.~A. Alves~Jr.\ {\em et~al.},
  \ifthenelse{\boolean{articletitles}}{\emph{{The \lhcb detector at the LHC}},
  }{}\href{http://dx.doi.org/10.1088/1748-0221/3/08/S08005}{JINST \textbf{3}
  (2008) S08005}\relax
\mciteBstWouldAddEndPuncttrue
\mciteSetBstMidEndSepPunct{\mcitedefaultmidpunct}
{\mcitedefaultendpunct}{\mcitedefaultseppunct}\relax
\EndOfBibitem
\bibitem{LHCb-DP-2012-003}
M.~Adinolfi {\em et~al.},
  \ifthenelse{\boolean{articletitles}}{\emph{{Performance of the \lhcb RICH
  detector at the LHC}},
  }{}\href{http://dx.doi.org/10.1140/epjc/s10052-013-2431-9}{Eur.\ Phys.\ J.\
  \textbf{C73} (2013) 2431}, \href{http://arxiv.org/abs/1211.6759}{{\tt
  arXiv:1211.6759}}\relax
\mciteBstWouldAddEndPuncttrue
\mciteSetBstMidEndSepPunct{\mcitedefaultmidpunct}
{\mcitedefaultendpunct}{\mcitedefaultseppunct}\relax
\EndOfBibitem
\bibitem{LHCb-DP-2012-002}
A.~A. Alves~Jr.\ {\em et~al.},
  \ifthenelse{\boolean{articletitles}}{\emph{{Performance of the LHCb muon
  system}}, }{}\href{http://dx.doi.org/10.1088/1748-0221/8/02/P02022}{JINST
  \textbf{8} (2013) P02022}, \href{http://arxiv.org/abs/1211.1346}{{\tt
  arXiv:1211.1346}}\relax
\mciteBstWouldAddEndPuncttrue
\mciteSetBstMidEndSepPunct{\mcitedefaultmidpunct}
{\mcitedefaultendpunct}{\mcitedefaultseppunct}\relax
\EndOfBibitem
\bibitem{LHCb-DP-2012-004}
R.~Aaij {\em et~al.}, \ifthenelse{\boolean{articletitles}}{\emph{{The \lhcb
  trigger and its performance in 2011}},
  }{}\href{http://dx.doi.org/10.1088/1748-0221/8/04/P04022}{JINST \textbf{8}
  (2013) P04022}, \href{http://arxiv.org/abs/1211.3055}{{\tt
  arXiv:1211.3055}}\relax
\mciteBstWouldAddEndPuncttrue
\mciteSetBstMidEndSepPunct{\mcitedefaultmidpunct}
{\mcitedefaultendpunct}{\mcitedefaultseppunct}\relax
\EndOfBibitem
\bibitem{BBDT}
V.~V. Gligorov and M.~Williams,
  \ifthenelse{\boolean{articletitles}}{\emph{{Efficient, reliable and fast
  high-level triggering using a bonsai boosted decision tree}},
  }{}\href{http://dx.doi.org/10.1088/1748-0221/8/02/P02013}{JINST \textbf{8}
  (2013) P02013}, \href{http://arxiv.org/abs/1210.6861}{{\tt
  arXiv:1210.6861}}\relax
\mciteBstWouldAddEndPuncttrue
\mciteSetBstMidEndSepPunct{\mcitedefaultmidpunct}
{\mcitedefaultendpunct}{\mcitedefaultseppunct}\relax
\EndOfBibitem
\bibitem{Sjostrand:2007gs}
T.~Sj\"{o}strand, S.~Mrenna, and P.~Skands,
  \ifthenelse{\boolean{articletitles}}{\emph{{A brief introduction to PYTHIA
  8.1}}, }{}\href{http://dx.doi.org/10.1016/j.cpc.2008.01.036}{Comput.\ Phys.\
  Commun.\  \textbf{178} (2008) 852},
  \href{http://arxiv.org/abs/0710.3820}{{\tt arXiv:0710.3820}}\relax
\mciteBstWouldAddEndPuncttrue
\mciteSetBstMidEndSepPunct{\mcitedefaultmidpunct}
{\mcitedefaultendpunct}{\mcitedefaultseppunct}\relax
\EndOfBibitem
\bibitem{LHCb-PROC-2010-056}
I.~Belyaev {\em et~al.}, \ifthenelse{\boolean{articletitles}}{\emph{{Handling
  of the generation of primary events in Gauss, the LHCb simulation
  framework}}, }{}\href{http://dx.doi.org/10.1088/1742-6596/331/3/032047}{{J.\
  Phys.\ Conf.\ Ser.\ } \textbf{331} (2011) 032047}\relax
\mciteBstWouldAddEndPuncttrue
\mciteSetBstMidEndSepPunct{\mcitedefaultmidpunct}
{\mcitedefaultendpunct}{\mcitedefaultseppunct}\relax
\EndOfBibitem
\bibitem{Lange:2001uf}
D.~J. Lange, \ifthenelse{\boolean{articletitles}}{\emph{{The EvtGen particle
  decay simulation package}},
  }{}\href{http://dx.doi.org/10.1016/S0168-9002(01)00089-4}{Nucl.\ Instrum.\
  Meth.\  \textbf{A462} (2001) 152}\relax
\mciteBstWouldAddEndPuncttrue
\mciteSetBstMidEndSepPunct{\mcitedefaultmidpunct}
{\mcitedefaultendpunct}{\mcitedefaultseppunct}\relax
\EndOfBibitem
\bibitem{Golonka:2005pn}
P.~Golonka and Z.~Was, \ifthenelse{\boolean{articletitles}}{\emph{{PHOTOS Monte
  Carlo: A precision tool for QED corrections in $Z$ and $W$ decays}},
  }{}\href{http://dx.doi.org/10.1140/epjc/s2005-02396-4}{Eur.\ Phys.\ J.\
  \textbf{C45} (2006) 97}, \href{http://arxiv.org/abs/hep-ph/0506026}{{\tt
  arXiv:hep-ph/0506026}}\relax
\mciteBstWouldAddEndPuncttrue
\mciteSetBstMidEndSepPunct{\mcitedefaultmidpunct}
{\mcitedefaultendpunct}{\mcitedefaultseppunct}\relax
\EndOfBibitem
\bibitem{Allison:2006ve}
Geant4 collaboration, J.~Allison {\em et~al.},
  \ifthenelse{\boolean{articletitles}}{\emph{{Geant4 developments and
  applications}}, }{}\href{http://dx.doi.org/10.1109/TNS.2006.869826}{IEEE
  Trans.\ Nucl.\ Sci.\  \textbf{53} (2006) 270}\relax
\mciteBstWouldAddEndPuncttrue
\mciteSetBstMidEndSepPunct{\mcitedefaultmidpunct}
{\mcitedefaultendpunct}{\mcitedefaultseppunct}\relax
\EndOfBibitem
\bibitem{Agostinelli:2002hh}
Geant4 collaboration, S.~Agostinelli {\em et~al.},
  \ifthenelse{\boolean{articletitles}}{\emph{{Geant4: A simulation toolkit}},
  }{}\href{http://dx.doi.org/10.1016/S0168-9002(03)01368-8}{Nucl.\ Instrum.\
  Meth.\  \textbf{A506} (2003) 250}\relax
\mciteBstWouldAddEndPuncttrue
\mciteSetBstMidEndSepPunct{\mcitedefaultmidpunct}
{\mcitedefaultendpunct}{\mcitedefaultseppunct}\relax
\EndOfBibitem
\bibitem{LHCb-PROC-2011-006}
M.~Clemencic {\em et~al.}, \ifthenelse{\boolean{articletitles}}{\emph{{The
  \lhcb simulation application, Gauss: Design, evolution and experience}},
  }{}\href{http://dx.doi.org/10.1088/1742-6596/331/3/032023}{{J.\ Phys.\ Conf.\
  Ser.\ } \textbf{331} (2011) 032023}\relax
\mciteBstWouldAddEndPuncttrue
\mciteSetBstMidEndSepPunct{\mcitedefaultmidpunct}
{\mcitedefaultendpunct}{\mcitedefaultseppunct}\relax
\EndOfBibitem
\bibitem{Hulsbergen:2005pu}
W.~D. Hulsbergen, \ifthenelse{\boolean{articletitles}}{\emph{{Decay chain
  fitting with a Kalman filter}},
  }{}\href{http://dx.doi.org/10.1016/j.nima.2005.06.078}{Nucl.\ Instrum.\
  Meth.\  \textbf{A552} (2005) 566},
  \href{http://arxiv.org/abs/physics/0503191}{{\tt
  arXiv:physics/0503191}}\relax
\mciteBstWouldAddEndPuncttrue
\mciteSetBstMidEndSepPunct{\mcitedefaultmidpunct}
{\mcitedefaultendpunct}{\mcitedefaultseppunct}\relax
\EndOfBibitem
\bibitem{Breiman}
L.~Breiman, J.~H. Friedman, R.~A. Olshen, and C.~J. Stone, {\em Classification
  and regression trees}, Wadsworth international group, Belmont, California,
  USA, 1984\relax
\mciteBstWouldAddEndPuncttrue
\mciteSetBstMidEndSepPunct{\mcitedefaultmidpunct}
{\mcitedefaultendpunct}{\mcitedefaultseppunct}\relax
\EndOfBibitem
\bibitem{AdaBoost}
R.~E. Schapire and Y.~Freund, \ifthenelse{\boolean{articletitles}}{\emph{A
  decision-theoretic generalization of on-line learning and an application to
  boosting}, }{}\href{http://dx.doi.org/10.1006/jcss.1997.1504}{Jour.\ Comp.\
  and Syst.\ Sc.\  \textbf{55} (1997) 119}\relax
\mciteBstWouldAddEndPuncttrue
\mciteSetBstMidEndSepPunct{\mcitedefaultmidpunct}
{\mcitedefaultendpunct}{\mcitedefaultseppunct}\relax
\EndOfBibitem
\bibitem{LHCb-CONF-2013-003}
{LHCb collaboration}, \ifthenelse{\boolean{articletitles}}{\emph{{A search for
  time-integrated \CP violation in $D^0 \to K^-K^+$ and $D^0 \to \pi^-\pi^+$
  decays}}, }{}
  \href{http://cdsweb.cern.ch/search?p=LHCb-CONF-2013-003&f=reportnumber&actio%
n_search=Search&c=LHCb+Conference+Contributions} {LHCb-CONF-2013-003}\relax
\mciteBstWouldAddEndPuncttrue
\mciteSetBstMidEndSepPunct{\mcitedefaultmidpunct}
{\mcitedefaultendpunct}{\mcitedefaultseppunct}\relax
\EndOfBibitem
\bibitem{Jackson:1964zd}
J.~D. Jackson, \ifthenelse{\boolean{articletitles}}{\emph{{Remarks on the
  phenomenological analysis of resonances}},
  }{}\href{http://dx.doi.org/10.1007/BF02750563}{Nuovo Cim.\  \textbf{34}
  (1964) 1644}\relax
\mciteBstWouldAddEndPuncttrue
\mciteSetBstMidEndSepPunct{\mcitedefaultmidpunct}
{\mcitedefaultendpunct}{\mcitedefaultseppunct}\relax
\EndOfBibitem
\bibitem{LHCb-DP-2013-002}
LHCb collaboration, R.~Aaij {\em et~al.},
  \ifthenelse{\boolean{articletitles}}{\emph{{Measurement of the track
  reconstruction efficiency at LHCb}},
  }{}\href{http://dx.doi.org/10.1088/1748-0221/10/02/P02007}{JINST \textbf{10}
  (2015) P02007}, \href{http://arxiv.org/abs/1408.1251}{{\tt
  arXiv:1408.1251}}\relax
\mciteBstWouldAddEndPuncttrue
\mciteSetBstMidEndSepPunct{\mcitedefaultmidpunct}
{\mcitedefaultendpunct}{\mcitedefaultseppunct}\relax
\EndOfBibitem
\bibitem{Wilks:1938dza}
S.~S. Wilks, \ifthenelse{\boolean{articletitles}}{\emph{{The large-sample
  distribution of the likelihood ratio for testing composite hypotheses}},
  }{}\href{http://dx.doi.org/10.1214/aoms/1177732360}{Annals Math.\ Statist.\
  \textbf{9} (1938) 60}\relax
\mciteBstWouldAddEndPuncttrue
\mciteSetBstMidEndSepPunct{\mcitedefaultmidpunct}
{\mcitedefaultendpunct}{\mcitedefaultseppunct}\relax
\EndOfBibitem
\bibitem{LHCb-PAPER-2011-037}
LHCb collaboration, R.~Aaij {\em et~al.},
  \ifthenelse{\boolean{articletitles}}{\emph{{Measurement of prompt hadron
  production ratios in $pp$ collisions at $\sqrt{s} = $ 0.9 and 7 TeV}},
  }{}\href{http://dx.doi.org/10.1140/epjc/s10052-012-2168-x}{Eur.\ Phys.\ J.\
  \textbf{C72} (2012) 2168}, \href{http://arxiv.org/abs/1206.5160}{{\tt
  arXiv:1206.5160}}\relax
\mciteBstWouldAddEndPuncttrue
\mciteSetBstMidEndSepPunct{\mcitedefaultmidpunct}
{\mcitedefaultendpunct}{\mcitedefaultseppunct}\relax
\EndOfBibitem
\bibitem{LHCb-PAPER-2012-041}
LHCb collaboration, R.~Aaij {\em et~al.},
  \ifthenelse{\boolean{articletitles}}{\emph{{Prompt charm production in $pp$
  collisions at $\sqrt{s}=7$ TeV}},
  }{}\href{http://dx.doi.org/10.1016/j.nuclphysb.2013.02.010}{Nucl.\ Phys.\
  \textbf{B871} (2013) 1}, \href{http://arxiv.org/abs/1302.2864}{{\tt
  arXiv:1302.2864}}\relax
\mciteBstWouldAddEndPuncttrue
\mciteSetBstMidEndSepPunct{\mcitedefaultmidpunct}
{\mcitedefaultendpunct}{\mcitedefaultseppunct}\relax
\EndOfBibitem
\bibitem{LHCb-PAPER-2012-037}
LHCb collaboration, R.~Aaij {\em et~al.},
  \ifthenelse{\boolean{articletitles}}{\emph{{Measurement of the fragmentation
  fraction ratio $f_s/f_d$ and its dependence on $B$ meson kinematics}},
  }{}\href{http://dx.doi.org/10.1007/JHEP04(2013)001}{JHEP \textbf{04} (2013)
  001}, \href{http://arxiv.org/abs/1301.5286}{{\tt arXiv:1301.5286}}\relax
\mciteBstWouldAddEndPuncttrue
\mciteSetBstMidEndSepPunct{\mcitedefaultmidpunct}
{\mcitedefaultendpunct}{\mcitedefaultseppunct}\relax
\EndOfBibitem
\bibitem{LHCb-PAPER-2015-041}
LHCb collaboration, R.~Aaij {\em et~al.},
  \ifthenelse{\boolean{articletitles}}{\emph{{Measurements of prompt charm
  production cross-sections in $pp$ collisions at $\sqrt{s} = 13\,$TeV}},
  }{}\href{http://arxiv.org/abs/1510.01707}{{\tt arXiv:1510.01707}}
  {LHCb-PAPER-2015-041}, \href{http://arxiv.org/abs/1510.01707}{{\tt
  arXiv:1510.01707}}, {submitted to JHEP}\relax
\mciteBstWouldAddEndPuncttrue
\mciteSetBstMidEndSepPunct{\mcitedefaultmidpunct}
{\mcitedefaultendpunct}{\mcitedefaultseppunct}\relax
\EndOfBibitem
\end{mcitethebibliography}

\ifx\mcitethebibliography\mciteundefinedmacro
\PackageError{LHCb.bst}{mciteplus.sty has not been loaded}
{This bibstyle requires the use of the mciteplus package.}\fi
\providecommand{\href}[2]{#2}

\newpage

\newpage

\centerline{\large\bf LHCb collaboration}
\begin{flushleft}
\small
R.~Aaij$^{38}$, 
B.~Adeva$^{37}$, 
M.~Adinolfi$^{46}$, 
A.~Affolder$^{52}$, 
Z.~Ajaltouni$^{5}$, 
S.~Akar$^{6}$, 
J.~Albrecht$^{9}$, 
F.~Alessio$^{38}$, 
M.~Alexander$^{51}$, 
S.~Ali$^{41}$
% ...

{\footnotesize \it
% ...
$ ^{5}$Clermont Universit\'{e}, Universit\'{e} Blaise Pascal, CNRS/IN2P3, LPC, Clermont-Ferrand, France\\
$ ^{6}$CPPM, Aix-Marseille Universit\'{e}, CNRS/IN2P3, Marseille, France\\
% ...
$ ^{9}$Fakult\"{a}t Physik, Technische Universit\"{a}t Dortmund, Dortmund, Germany\\
% ...
$ ^{37}$Universidad de Santiago de Compostela, Santiago de Compostela, Spain\\
$ ^{38}$European Organization for Nuclear Research (CERN), Geneva, Switzerland\\
% ...
$ ^{41}$Nikhef National Institute for Subatomic Physics, Amsterdam, The Netherlands\\
% ...
$ ^{46}$H.H. Wills Physics Laboratory, University of Bristol, Bristol, United Kingdom\\
% ...
$ ^{51}$School of Physics and Astronomy, University of Glasgow, Glasgow, United Kingdom\\
$ ^{52}$Oliver Lodge Laboratory, University of Liverpool, Liverpool, United Kingdom\\
% ...
}
\end{flushleft}

\end{document}